\documentclass[12pt]{article}
\usepackage{epsf}
\usepackage{multirow}
\usepackage{graphicx}

\begin{document}


\begin{center}
{\bf Precision Measurements of Electroweak Parameters with Z Bosons at the Tevatron}\\
 \ \\
 Arie  Bodek  for the CDF and D0 collaborations\\
{\em Department of Physics and Astronomy,\\ University of Rochester, Rochester, NY. 14627, USA} \\
Proceedings of the Third Annual Large Hadron Collider Physics Conference\\
 LHCP15, Aug. 31-Sept 5, 2015   Saint Petersburg, Russia

\vspace*{1.0cm}
ABSTRACT \\
\end{center}
\noindent
We report on the extraction of  $\sin^2\theta^{\rm lept}_{\rm eff}(M_Z)$  and an  indirect measurement of
the mass of the W boson from the forward-backward asymmetry of dilepton 
events in the $Z$ boson  mass region at the Tevatron.  The data samples
of $e^+e^-$ and $\mu^+\mu^-$ events collected by the CDF detector
correspond to the full 9.4 fb$^{-1}$  run II sample and 
yield an effective electroweak mixing angle   $\sin^2\theta^{\rm lept}_{\rm eff}(M_Z) =  0.23222 \pm 0.00046$.  The corresponding result reported by the D0 collaboration with  the full 9.4 fb$^{-1}$  $e^+e-$ sample
is  $\sin^2\theta^{\rm lept}_{\rm eff}(M_Z) =  0.23146 \pm 0.00047$.
The CDF collaboration also extracts the on-shell electroweak mixing angle
 $ \sin^2 \theta_W  =  0.22401 \pm 0.00044$ which corresponds to
 an indirect measurement of the W boson mass   
  $M_W ({\rm indirect})  =  80.327 \pm 0.023 \;{\rm GeV}$.
The quoted uncertainties include both statistical and systematic
contributions.

\section{Introduction}
The effective $\sin^2 \theta_W$ coupling at the lepton vertex, denoted as  $\sin^2\theta^{\rm lept}_{\rm eff}(M_Z)$, has been
accurately measured at the LEP-1 and SLD $e^+e^-$ colliders. The combined average of six individual  LEP-1 and SLD  
measurements\cite{LEPfinalZ} yields  $\sin^2\theta^{\rm lept}_{\rm eff}(M_Z) = 0.23153 \pm 0.00016$. However, there is tension between the two most precise individual measurements: the combined LEP-1 and SLD $b$-quark forward-backward asymmetry ($A_{\rm FB}^{0,{\rm b}})$ yields  $\sin^2\theta^{\rm lept}_{\rm eff}(M_Z) = 0.23221 \pm 0.00029$,
and the SLD polarized left-right asymmetry  $({\cal A}_\ell)$ yields  $\sin^2\theta^{\rm lept}_{\rm eff}(M_Z) = 0.23098 \pm 0.00026$. These two measurements differ by 3.2 standard deviations.
In order to help resolve this difference
 new measurements  of   $\sin^2\theta^{\rm lept}_{\rm eff}(M_Z)$ 
 should have uncertainties similar to SLD or  LEP ($\approx \pm$0.0003).

\begin{figure}[ht]
 \centering
\includegraphics[width=6.in,height=2.5in]{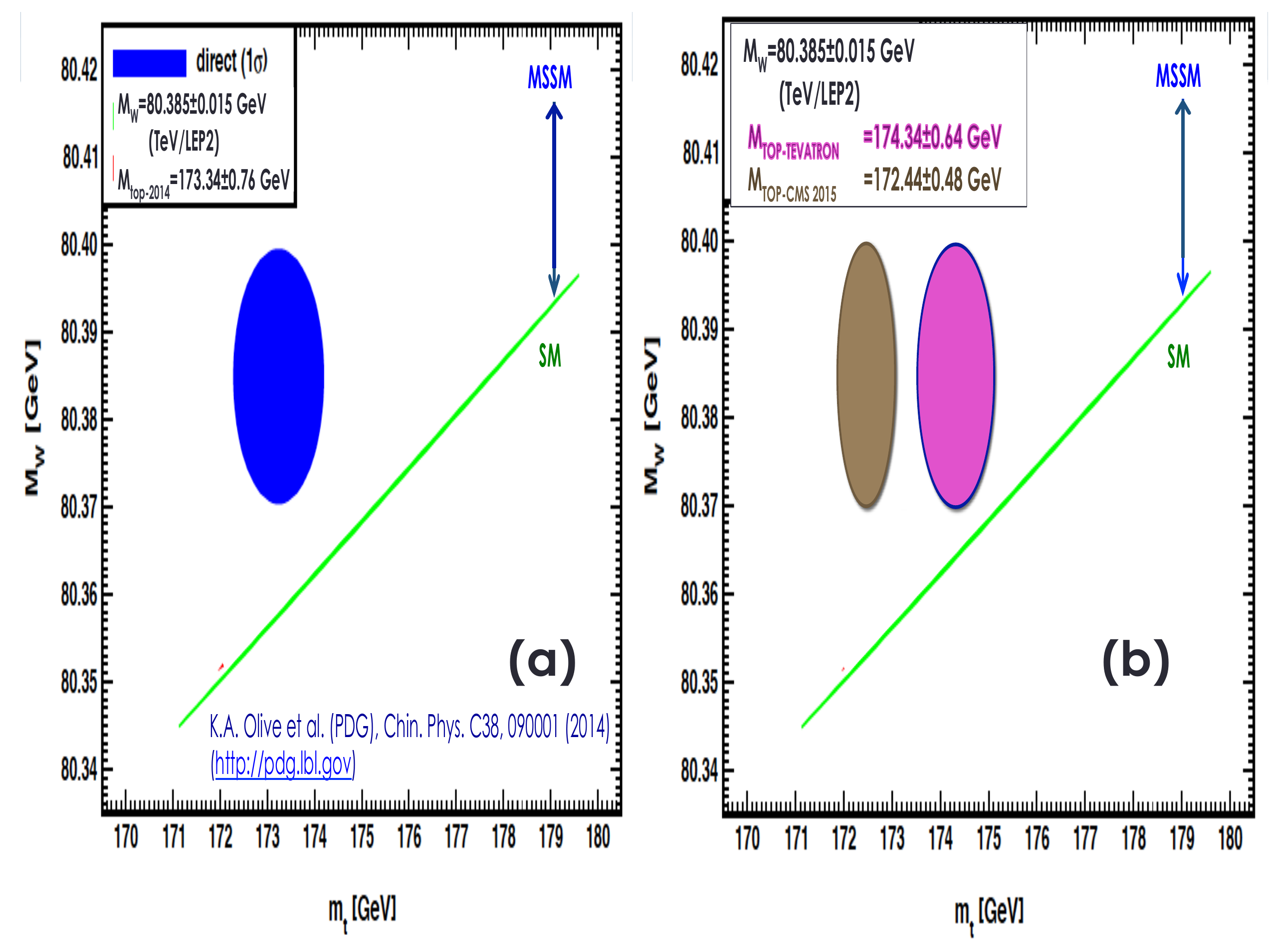}
\caption{ (a) World average of all direct measurements of $M_W$  (CDF, D0, LEP2) versus 
the average of all  $M_t$ measurements  (CDF, D0, CMS, ATLAS) in  2014.  The green line is the expectation from the SM (with $M_H=125.6\pm0.7$ GeV).  Supersymmetry models predict values which are above the SM line.  (b) Same as (a) but with the CMS measurement of  $M_t$ in 2015 as compared to the Tevatron
measurement of $M_t$.
 }
\label{Fig1}
\end{figure}

In addition, now that the Higgs boson mass  ($M_H$) is known, the Standard Model  (SM) is over constrained. Any inconsistency between precise measurements  of SM parameters could be indicative of new physics.  
 Fig.\ref{Fig1} (a)  (from ref.\cite{PDG}) shows the current world average\cite{w2014}  of direct measurements of the mass
 of the W boson  ($M_W$=$80.385\pm0.015$ GeV)  versus the 2014  average\cite{top2014} of the direct measurements of
 the mass of the top quark ($M_t =173.34\pm0.76$ GeV).     

 The average of the Tevatron measurements of $M_t$ in 2014 is $M_t$=174.34 $\pm$ 0.37(stat)  $\pm$ 0.52(syst)  GeV (or 174.34$\pm$0.64).  If we also include  the 2014 measurements of ATLAS and CMS the combined  2014 world average \cite{top2014}  (CDF, D0, CMS, ATLAS) is $M_t$=173.34 $\pm$ 0.27(stat) $\pm$ 0.71(syst)  GeV (or 173.34 $\pm$ 0.76 GeV) as shown in   Fig.\ref{Fig1} (a).  Also shown in green is the expectation from the SM with $M_H=125.6\pm0.7$ GeV.   The average of  all direct measurements of $M_W$   is about 1.5 standard deviation higher  than the prediction of the standard
    model.  Predictions of  supersymmetric models
     for $M_W$ are also higher \cite{super} than the predictions of the standard model.

      The most recent measurement  of $M_t$ at the LHC are somewhat lower than at the Tevatron. 
  The  ATLAS\cite{ATLAStop} measurement published in  2015 is $M_t$= 172.99 $\pm$ 0.91 GeV.   The CMS\cite{CMStop} 2015 measurement   $M_t$=172.44 $\pm$ 0.13(stat) $\pm$ 0.47(syst)  GeV (or 172.44 $\pm$0.48 GeV) is the most precise measurement to date  and supersedes all previous CMS results.  There is about a two standard deviation tension between the 2015 CMS measurement of $M_t$  and  the earlier Tevatron measurements.  However,  both are consistent with the world average.   The lower value of $M_t$  as 
    %
   measured by  CMS would imply a somewhat larger deviation of $M_W$ from the prediction of the SM as shown in Fig.~\ref{Fig1} (b).
   The parameter that needs to be measured more precisely is  $M_W$. The current experimental uncertainties in the  direct measurements of the W boson mass ($M_W^{direct}$)  by D0 and CDF at the Tevatron are about  $\pm$20  MeV per experiment.  Equivalently  one can also measure the on-shell\cite{onshell} weak mixing angle,  $\sin^2\theta_W= 1 - M_W^2/M_Z^2$.   An error of $\pm$ 0.0004  in  the on-shell  $\sin^2\theta_W$ is equivalent to an indirect measurement
 of  the W boson mass  ($M_W^{indirect}$)  to a precision of $\pm$ 20 MeV. 
 
The angular distribution for the production of deletions in hadron colliders is proportional to
 $$1+\cos^2\theta+\frac{A_0}{2}(1-3\cos^2\theta)+A_4\cos\theta,$$
 where $\theta$ is the polar angle in the Collins-Soper frame\cite{collins}.  The coefficient
$A_0(P_T)$ is small and vanished for dilepton transverse momentum $P_T=0$.  The integrated forward-backward asymmetry
$A_{\rm fb}(M)$ is equal to  $3A_4(M)/8$,

Precise extractions of  $\sin^2\theta^{\rm lept}_{\rm eff}(M_Z)$  and  $\sin^2\theta_W= 1 - M_W^2/M_Z^2$  using the
forward-backward asymmetry  ($A_{\rm fb}$) of dilepton events produced in 
$p\bar p$ and $pp$ collisions  are now  possible for the first time because of  four new innovations:
\begin{itemize}
\item A new technique \cite{muon-scale} for calibrating the muon and electron energy scales
as a function of  detector  $\eta$ and $\phi$ (and sign), thus greatly reducing systematic            
uncertainties from the energy scale.   These technique is used at CDF and CMS.  A similar technique is used
by D0 for electrons.
\item A new  event weighting technique\cite{event-weighting}.  With this technique
all  experimental uncertainties in acceptance and efficiencies cancel (by measuring  the $\cos\theta$ coefficient $A_4$  and  using the relation
 $A_{\rm fb}=3A_4/8$). Similarly,  additional weights can be  included for antiquark dilution, which makes the analysis independent of the acceptance in dilepton rapidity.  These technique is used by CDF and is currently being implemented at CMS.
\item The  implementation\cite{cdf-ee} in 2012 of Z fitter Effective Born Approximation (EBA) electroweak radiative corrections into the theory modified predictions of \textsc{powheg} and \textsc{resbos} which 
 allows for a measurement of both  $\sin^2\theta^{\rm lept}_{\rm eff}$(M$_Z$) and 
  $\sin^2\theta_W= 1 - M_W^2/M_Z^2$.  These EBA  electroweak radiative corrections were implemented in  CDF analyses\cite{cdf-ee,cdf-mumu, CDFweb} since  2013.
  Recently, an official version of  \textsc{powheg} with electroweak radiative corrections has been released.  Similarly, electroweak radiative corrections have been implemented in other theory predictions. Comparisons of different  implementation of EW radiative corrections  are now possible..

  \item A new technique \cite{pdf_error} that reduces Parton Distribution Function (PDF) uncertainties
  by incorporating additional constraints from the mass and rapidity dependence of Drell-Yan $A_{\rm fb}$.
  The use of Drell-Yan  $A_{\rm fb}(M,y)$  $\chi^2$ weighting was  first proposed in ref. \cite{pdf_error}) for additional constraints on PDFs.  The $\chi^2$ weighting  technique reduces the PDF uncertainty in the   measurements of 
  $\sin^2\theta^{\rm lept}_{\rm eff}$(M$_Z$),  $\sin^2\theta_W$, and  in the indirect and
  direct measurements of   $M_W$.  This technique has been used in CDF\cite{CDFweb} and is currently being implemented in CMS.
\end{itemize}

\subsection{Momentum-energy scale corrections}

This new technique\cite{muon-scale} is used in CDF (for both muons and electrons) and also  in CMS. In  CMS  it is used to get a precise measurement of the  Higgs boson mass in the four lepton channel.  A similar technique is used
by D0 for electrons. The technique used in CDF and CMS  relies on the fact that the $Z$ boson mass is well known as follows:
\begin{itemize}   
\item  Any  correlation between the scales of the two leptons is removed by getting an initial calibration using  Z events. It is done by requiring that the  mean  $\langle{1/P_T} \rangle$  of each lepton in bins of  detector $\eta$, $\phi$   and charge is equal to the expected value for generated Z events, smeared by the momentum/energy resolution.
\item    The Z boson mass is is used as a second order correction.  The measured  Z boson mass as a function of detector  $\eta$, $\phi$ and charge  of the lepton is required to be equal to the value for generated Z events (smeared by the momentum/energy resolution).  Additionally the  measured  J/$\psi$ and $\Upsilon$ masses as a function of $\eta$ of the lepton are also used.
\end{itemize}
The scale corrections are determined for both data events and reconstructed  hit level Monte Carlo events.
After corrections, the  reconstructed Z boson mass as a function $\eta$, $\phi$ and charge  for both the data and hit level  MC agrees with the generator level Monte Carlo (smeared by resolution, and with experimental acceptance cuts). All charge bias is removed.  For muons, the following calibration constants are extracted for each  bin in  $\eta$ and $\phi$
\begin{itemize}  
\item  A multiplicative calibration correction in the quantity $1/P_T$  which accounts for possible mis-calibration of the magnetic field.
\item A calibration correction which is additive in $1/P_T$  which  accounts for tracker mis-alignments.  
\item For very low energy muons,  the J/$\psi$ and $\Upsilon$ masses are used to determine a small  additional calibration constant to tune the  dE/dx energy loss in the amount of material in the tracker as a function of  detector $\eta$.
\end{itemize}
 When the technique is used for electrons, the multiplicative correction accounts for tower mis-calibration and there is no additive correction since the tracker is not used in the reconstruction of the electron energy. 
\subsection{The event weighting technique}
The forward-backward  $A_{\rm fb}$  asymmetry of leptons measured with this technique\cite{event-weighting} is insensitive to  the acceptance and lepton detection efficiency. Therefore, the raw  $A_{\rm fb}$ which is measured using this technique is  automatically corrected for efficiency
and acceptance. The only corrections that need to be made are corrections for momentum/energy resolution which lead to event migration between different  bins in dilepton mass.  All  experiment dependent systematic uncertainties cancel to first order.  This technique is used in the CDF analysis for muons and electrons, and is currently being implemented at CMS.

The event weighting technique utilizes two kinds of weights.  Angular weights are used to remove the sensitivity to acceptance and lepton detection efficiency as a function of $\cos\theta$.  In the CDF (and CMS) analyses, only angular weights are used.  For proton-proton collisions at the LHC, one can also include weights which correct for the rapidity dependent dilution and therefore removes the sensitivity to the acceptance in dilepton rapidity.
\begin{figure}[h]
 \centering
\includegraphics[width=6.in,height=3.3in]{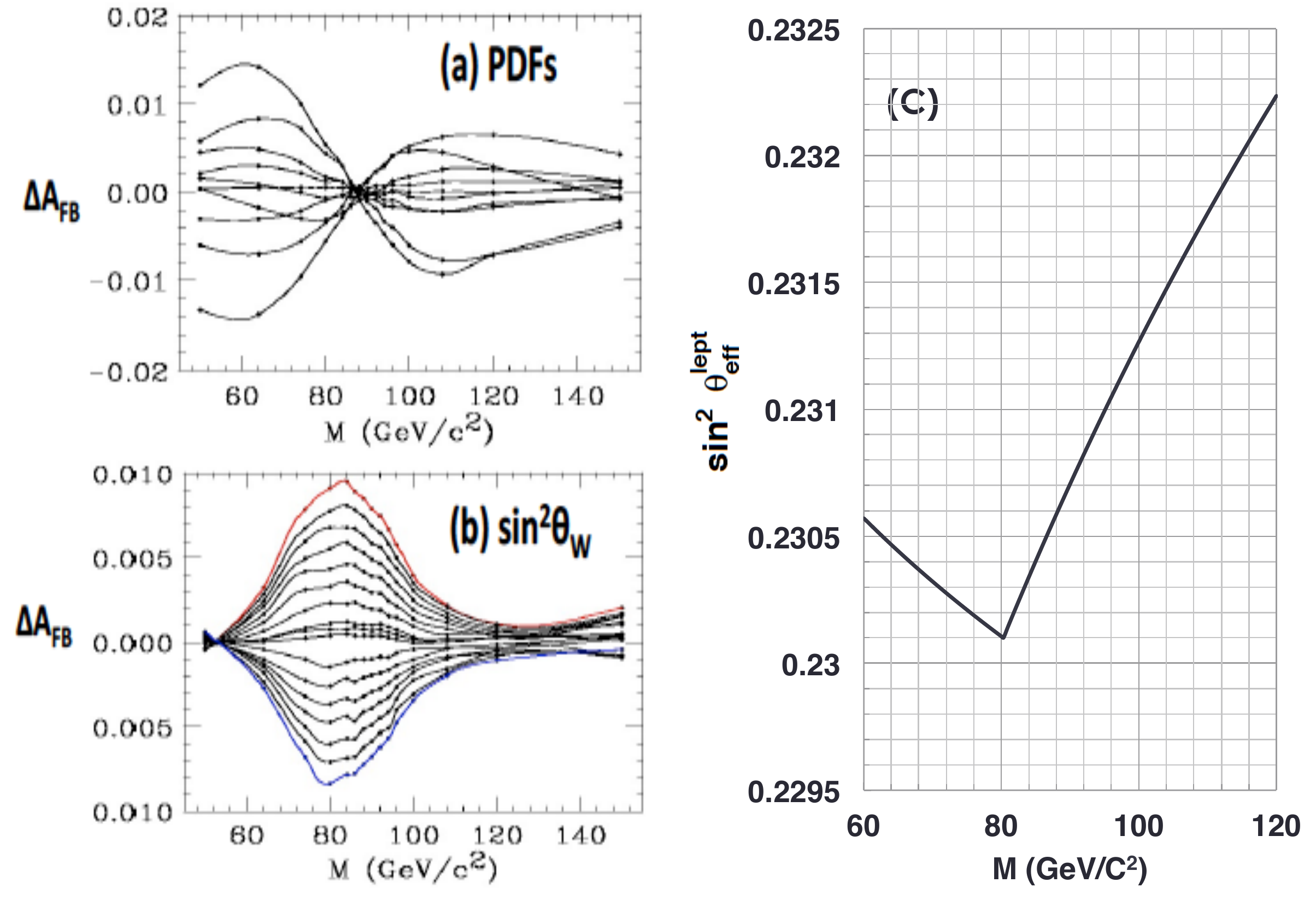}
\caption{ Tevatron: (a) The  difference between $A_{\rm fb}(M)$ 
for 10 \textsc{nnpdf3.0} (\textsc{nnlo}) replicas and $A_{\rm fb}(M)$ calculated
for  the  default \textsc{nnpdf3.0} (\textsc{nnlo}) (261000). Much of the difference
originates form the  different dilution factors for each of the \textsc{nnpdf}  replicas.
Here $\sin^2\theta_W$ is fixed at a value of 0.2244. (b) The  difference between 
$A_{\rm fb}(M)$  for different values of $\sin^2\theta_W$ ranging
from 0.2220 (shown at the top in red)  to 0.2265 (shown on the
bottom in blue), and 
$A_{\rm fb}(M)$ for $\sin^2\theta_W$=0.2244.
 Here  $A_{\rm fb}(M)$ is  calculated with  the default \textsc{nnpdf3.0} (\textsc{nnlo}). (Figures (a) and (b) are  from Ref. \cite{pdf_error}).   
(c) Scale dependence of $\sin^2\theta^{\rm lept}_{\rm eff}(M)$.  The minimum of  $\sin^2\theta^{\rm lept}_{\rm eff}(M)$ is
at the mass of the W boson (from Ref. \cite{PDG}).  }
\label{scale}
\end{figure}
\subsection {Electroweak radiative corrections}
\subsubsection{\textsc{zgrad}-type EW radiative corrections - used by D0}
An  approximate method that only corrects for the flavor dependence of  $\sin^2\theta_{\rm eff}$  has been proposed by Baur and collaborators \cite{baur}. The flavor dependence is approximately:
 $\sin^2\theta^{\rm u-quark}_{\rm eff} =    sin^2\theta^{\rm lept}_{\rm eff}- 0.0001$ and 
  $\sin^2\theta^{\rm d-quark}_{\rm eff} =    sin^2\theta^{\rm lept}_{\rm eff}- 0.0002$.

We refer to these EW corrections (which have been implemented in \textsc{resbos})  as \textsc{zgrad}-type corrections.  
These  corrections are used by D0. The D0 collaboration reports\cite{dzero} that 
  $\sin^2\theta^{\rm lept}_{\rm eff}$(M$_Z)$ extracted using \textsc{resbos} (with CTEQ 6.6 -\textsc{nlo} PDFs)
   including  \textsc{zgrad}-type radiative corrections is +0.00008 larger than the value of   $\sin^2\theta^{\rm lept}_{\rm eff}$(M$_Z)$ extracted using  \textsc{pythia} 6.323 \cite{pythia3} with
  the same PDF set and no EW radiative corrections. The   \textsc{pythia} matrix elements are QCD leading order as compared to  \textsc{resbos} matrix elements which are \textsc{nlo}. However, as reported by D0, the  estimated
correction due to higher order QCD effects is negligibly small.

The above procedure  partially corrects for the flavor dependence of $\sin^2\theta_{\rm eff}$. It does not account for the mass dependence of $\sin^2\theta_{\rm eff}$ (shown in Fig. \ref{scale}(c)) nor does it account for the complex mass dependent form
factors.   As described below, a  more complete treatment of EW radiative corrections factors is needed in order yield a measurement of   the on-shell $\sin^2\theta_W= 1 - M_W^2/M_Z^2$.

\subsubsection{Effective Born approximation (EBA) electroweak radiative corrections - used by CDF}
These radiative corrections have been implemented in  CDF\cite{cdf-ee} (for modified versions of  \textsc{powheg}, \textsc{resbos} and Tree level calculations). The corrections are  derived from the approach adopted at LEP\cite{fitter}.
The Z-scattering-amplitude form factors are calculated by ZFITTER 6.43 \cite{fitter} which  has been used by LEP-1 and SLD measurements for precision tests of the standard model \cite{sld-lep}.

$A_{fb}(M)$ in the region of the mass of the $Z$ boson is  sensitive to the effective weak mixing angle
 $\sin^2\theta_{\rm eff} (M, flavor)$, where $M$ is the dilepton mass.
  Here, $\sin^2\theta_{\rm eff}$ is related to the 
on-shell\cite{onshell}  electroweak mixing angle $\sin^2\theta_W= 1 - M_W^2/M_Z^2$ via complex mass and flavor (weak isospin) dependent  electroweak radiative corrections form factors. The massless-fermion approximation is used.

 The parameter which is measured at LEP and SLD is  $\sin^2\theta^{\rm lept}_{\rm eff}$(M$_Z)$.  Previous extraction of  
 $\sin^2\theta^{\rm lept}_{\rm eff}$(M$_Z$) from Drell-Yan  $A_{fb}$ neglected the dependence of   $\sin^2\theta_{\rm eff}$ on flavor
and dilepton mass.  The input to the theory predictions has been one value of   $\sin^2\theta_{\rm eff}$  which 
on average was  assumed to be independent of mass or flavor and has been interpreted as $\sin^2\theta^{\rm lept}_{\rm eff}$(M$_Z)$. 

When the full EBA  EW radiative corrections are included, the input to the theory prediction templates for 
 $A_{\rm fb}(M)$ is the on-shell  $\sin^2\theta_W= 1 - M_W^2/M_Z^2$. The templates
are compared to the data and the best fit value of $\sin^2\theta_W$ is extracted.
 From the best fit value of $\sin^2\theta_W$  and the full complex EBA radiative corrections form factors
we can then extract $\sin^2\theta^{\rm lept}_{\rm eff}(M_Z)$ which is the  effective leptonic
EW mixing angle at the mass of the Z boson. 
  With  the EBA radiative corrections used at CDF it is found that 
$\sin^2\theta^{\rm lept}_{\rm eff}(M_Z)\approx  1.037 \sin^2\theta_W.$

If the EBA EW radiative corrections are included, the extracted value of  $\sin^2\theta^{\rm lept}_{\rm eff}(M_Z)$
is higher by  +0.00023 than the value extracted with no EW radiative corrections. About +0.00008 originate from
accounting for the flavor dependence of   $\sin^2\theta^{\rm lept}_{\rm eff}(M)$, +0.00006 originates from
accounting for the mass dependence of $\sin^2\theta^{\rm lept}_{\rm eff}(M)$, and +0.00009 originate
from accounting for the mass dependent complex EW Fitter form factors.
%
%
%
\begin{figure}[h]
 \centering
\includegraphics[width=6.in,height=4.3in]{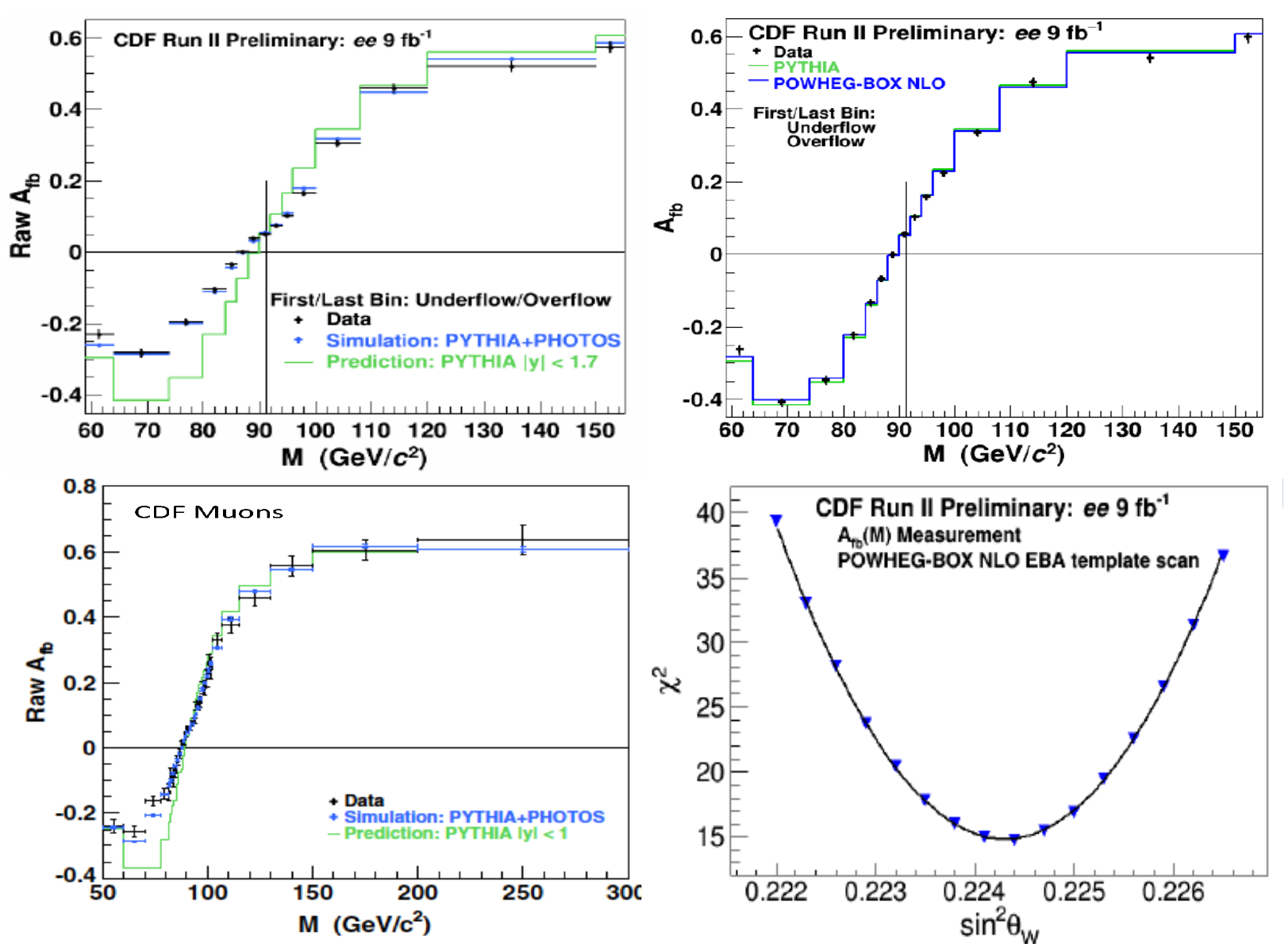}
\caption{ Top-Left:  CDF raw $A_{\rm fb}(M)$ measurement in bins of  $e^+e^-$
invariant mass.  Only statistical uncertainties (bin-by-bin unfolding) are shown.
The Monte Carlo
simulation (\textsc{pythia}) includes the effect of resolution smearing and FSR.
The \textsc{pythia} $|y|<1.7$ asymmetry curve does not.
Top-Right:  $A_{\rm fb}(M)$ for $e^+e^-$ events unfolded for resolution and QED-FSR.  The \textsc{pythia} calculation uses
$\sin^2\theta^{\rm lept}_{\rm eff} = 0.232$.
The EBA-based \textsc{resbos} and \textsc{powheg} calculations uses
$\sin^2\theta_W = 0.2233$ 
( $\sin^2\theta^{\rm lept}_{\rm eff}(M_Z) = 0.2315)$. 
Bottom-Left:  Same as Top-Left for the 
$\mu^+\mu^-$
 sample (here $|y|<1)$. Bottom-Right:   $\chi^2$  vs. $\sin^2\theta_W$
for the  CDF  $e^+e^-$ sample.}
\label{fig-raw-afb}
\end{figure}
\section{Analysis of CDF $\mu^+\mu^-$ and $e^+e^-$ full 9.4 fb$^{-1}$  run II sample}
%
After applying the calibrations and muon  and electron scale corrections to the experimental
and simulated data, $A_{\rm fb}(M)$ is measured in bins
of $\mu^+\mu^-$\cite{cdf-mumu} for  and $e^+e^-$\cite{CDFweb}  invariant mass using the event-weighting method.
This measurement is denoted as the raw $A_{\rm fb}(M)$
measurement because the event-weighting method provides a first-order
acceptance correction, but does not include resolution unfolding and
final-state (FSR) QED radiation.
The raw $A_{\rm fb}$ measurements in bins of the $\mu^+\mu^-$ and $e^+e^-$
invariant mass  are shown on the left part of  Fig.~\ref{fig-raw-afb}.  Only statistical uncertainties are shown.
The Monte Carlo
simulation (\textsc{pythia}+\textsc{photos}) includes the effect of resolution smearing and FSR.
To illustrate the effects of resolution smearing and FSR, the \textsc{pythia} $|y|<1$  and $|y|<1.7$  asymmetry curves do not include the effect of resolution smearing or  FSR.

With  the event weighting technique, the events near  $\cos\theta$=0 are assigned zero weight. Therefore,
the migration of events between positive and negative $\cos\theta$ is negligible. Resolution smearing
and FSR primarily transfer events between bins in invariant mass.
The raw $A_{\rm fb}$ in bins of $e^+e^-$ and $\mu^+\mu^-$ invariant mass is unfolded\cite{cdf-mumu} for resolution smearing and FSR using a transfer matrix
which is obtained from the Monte Carlo simulation. The unfolded $A_{\rm fb}(M)$ for electrons  is shown in the top-right panel of Fig. \ref{fig-raw-afb}.

The electroweak (EWK) mixing parameters $\sin^2\theta^{\rm lept}_{\rm eff}(M_Z)$ and
$\sin^2\theta_W$ are extracted from the 
 fully unfolded 
$A_{\rm fb}(M)$ measurements using
$A_{\rm fb}(M)$ templates calculated with different values of
$\sin^2\theta_W$. Three  QCD 
calculations are used: LO (tree), \textsc{resbos} \textsc{nlo}, and
\textsc{powheg-box} \textsc{nlo}.  The three calculations were modified
to include EWK radiative correction\cite{cdf-ee} using the
Effective Born Approximation (EBA).
%

The $A_{\rm fb}(M)$ measurement is directly sensitive to the
effective-mixing parameters $\sin^2\theta^{\rm lept}_{\rm eff}(M)$ which
are combinations of the form-factors and $\sin^2\theta_W$.  Most of
the sensitivity to   $\sin^2\theta^{\rm lept}_{\rm eff}(M_Z)$  comes from the Drell-Yan $A_{\rm fb}(M)$
near the Z pole, where  $A_{\rm fb}$ is small.  In contrast,  $A_{\rm fb}(M)$ at higher mass values  where  $A_{\rm fb}$
is large, is mostly sensitive to the axial coupling, which is known.
While the extracted values of the effective-mixing parameter  $\sin^2\theta^{\rm lept}_{\rm eff}(M_Z)$
are independent of the details of the EBA model, the
interpretation of the best-fit value of the on-shell $\sin^2\theta_W$ and its
corresponding form factors depend on the details
of the EBA model.

\begin{figure}[h]
 \centering
\includegraphics[width=6.in,height=4.0in]{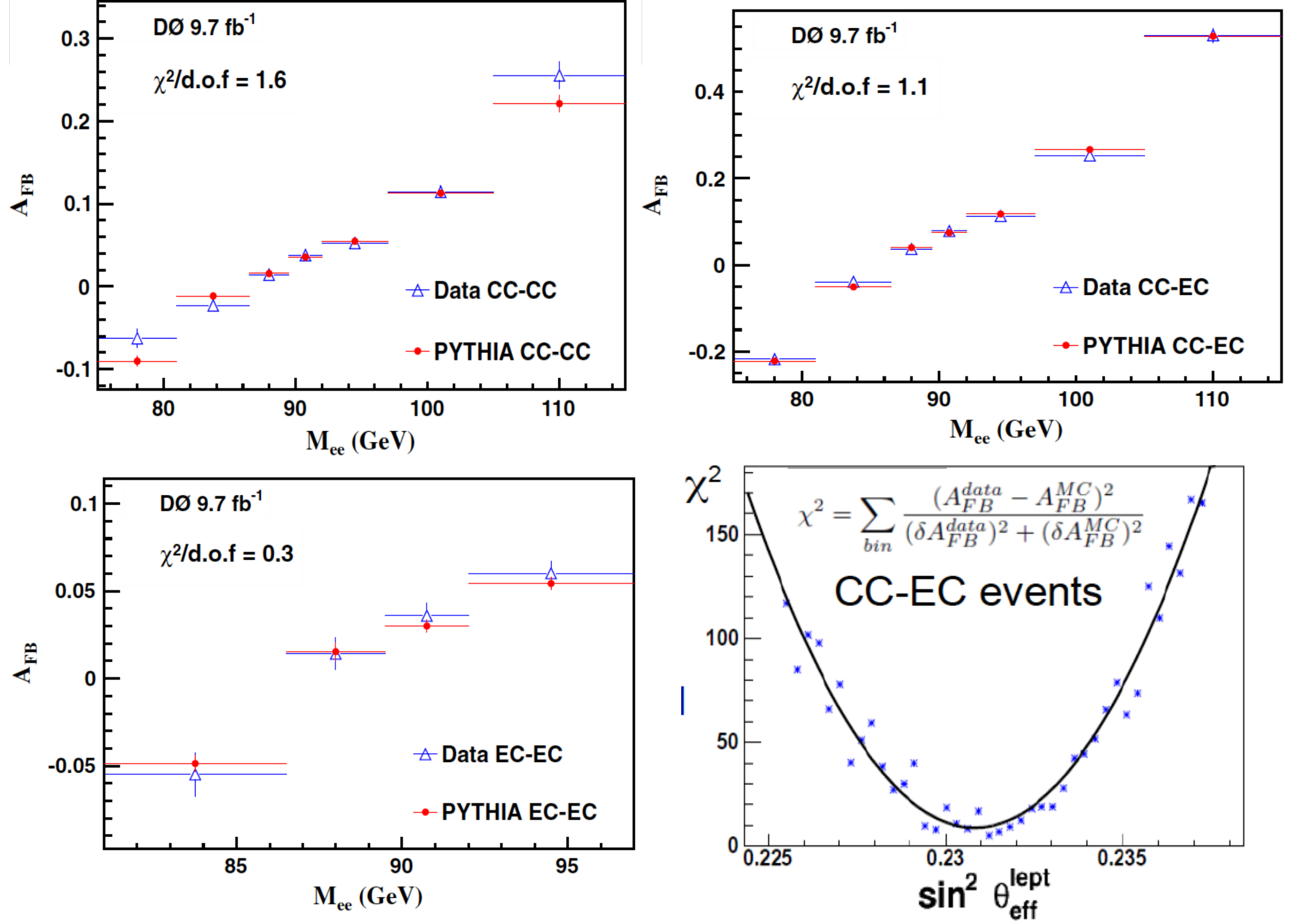}
\caption{ D0 raw $A_{\rm fb}(M)$ measurement in bins of $e^+e^-$
invariant mass for Central-Central calorimeters (CC-CC), Central-Endcap calorimeters (CC-CE), and Endcap-Endcap calorimeters (EC-EC) event topologies. Also shown is the $\chi^2$  vs. $\sin^2\theta^{\rm lept}_{\rm eff}$(M$_Z$)
for the    D0 $A_{\rm fb}(M)$ CC-CE topology. }
\label{fig-dzero-afb}
\end{figure}

Calculations of the $A_{\rm fb}(M)$ templates with different values of
the electroweak-mixing parameter are compared with the
measurement to determine the value of the parameter that
best describes the data. The calculations include both
quantum chromodynamic and EBA electroweak radiative corrections. 
The measurement and templates are compared using the $\chi^2$
statistic evaluated with the $A_{\rm fb}$ measurement
error matrix. Each template provides a scan point for the $\chi^2$ function
$(\sin^2\theta_W, \chi^2( \sin^2\theta_W))$. The scan points
are fit to a parabolic $\chi^2$ functional form.
 For the 
CDF $e^+e^-$ analysis, 
the $\chi^2$ distribution of the scan over templates from the
\textsc{powheg} \textsc{nlo} calculation (with \textsc{nnpdf3.0}) is shown in  
the bottom right panel of  Fig.~\ref{fig-raw-afb}.
For the  $e^+e^-$ analysis  the EBA-based \textsc{powheg} box NLO  \textsc{nnpdf3.0} calculations of $A_{\rm fb}(M)$
are used to extract the central value of $\sin^2\theta_W$.
For the CDF $\mu^+\mu^-$ analysis the EBA-based \textsc{resbos} (\textsc{cteq6.6m}) NLO  calculations of $A_{\rm fb}(M)$
are used to extract the central value of $\sin^2\theta_W$. The other
calculations are used to estimate the systematic uncertainty from
the electroweak radiative corrections and QCD \textsc{nlo} radiation.
\section{Analysis of D0  $e^+e^-$ full 9.4  fb$^{-1}$  run II sample}
In the published D0 analysis\cite{dzero},  $A_{\rm fb}(M)$ measurements in bins of $e^+e^-$ 
invariant mass are done for several event topologies as shown in Fig.\ref{fig-dzero-afb}.
Electrons and positrons are detected in 
in the Central Calorimeter (CC) and in the Endcap Calorimeter (EC). 
The event topologies correspond to Central-Central (CC-CC), Central-Endcap (CC-CE), and Endcap-Endcap  (EC-EC).  The effects of acceptance,  FSR and resolution smearing are all incorporated into MC templates with
different values of $\sin^2\theta^{\rm lept}$. The \textsc{resbos} templates (calculated
with \textsc{nnpdf2.3} \textsc{nlo} PDFs)  are compared to the data for the three topologies
and the  best fit values of  $\sin^2\theta^{\rm lept}$ are extracted. 
The  $\chi^2$  vs. $\sin^2\theta^{\rm lept}_{\rm eff}$(M$_Z$)
for the   D0 $A_{\rm fb}(M)$ CC-CE topology is shown in the bottom right panel of Fig.\ref{fig-dzero-afb}.
\subsection {Constraining PDFs through $\chi^2$ weighting}
 This technique which  was  first proposed in ref. \cite{pdf_error} has  been implemented in the most recent CDF analysis\cite{CDFweb}. At the Tevatron the technique 
 reduces the PDF uncertainty  in $\sin^2\theta_W$ by 20\%.    The reduction of the PDF uncertainty in $\sin^2\theta_W$ with this technique at the LHC is much more significant\cite{pdf_error}.
Fig. \ref{scale} (a) from Ref.\cite{pdf_error} shows the  difference between $A_{\rm fb}(M)$ 
for 10 \textsc{nnpdf3.0} (\textsc{nnlo}) replicas and $A_{\rm fb}(M)$ calculated
for  the  default \textsc{nnpdf3.0} (\textsc{nnlo}) (261000). Much of the difference
originates form the  different dilution factors for each of the \textsc{nnpdf}  replicas.
Here $\sin^2\theta_W$ is fixed at a value of 0.2244. Fig.\ref{scale}(b) 
shows the   difference between 
$A_{\rm fb}(M)$  for different values of $\sin^2\theta_W$ ranging
from 0.2220 (shown at the top in red)  to 0.2265 (shown on the
bottom in blue), and  $A_{\rm fb}(M)$ for $\sin^2\theta_W$=0.2244.
 Here  $A_{\rm fb}(M)$ is  calculated with  the default \textsc{nnpdf3.0} (\textsc{nnlo}).
   \begin{figure}[h]
 \centering
\includegraphics[width=6.in,height=3.5in]{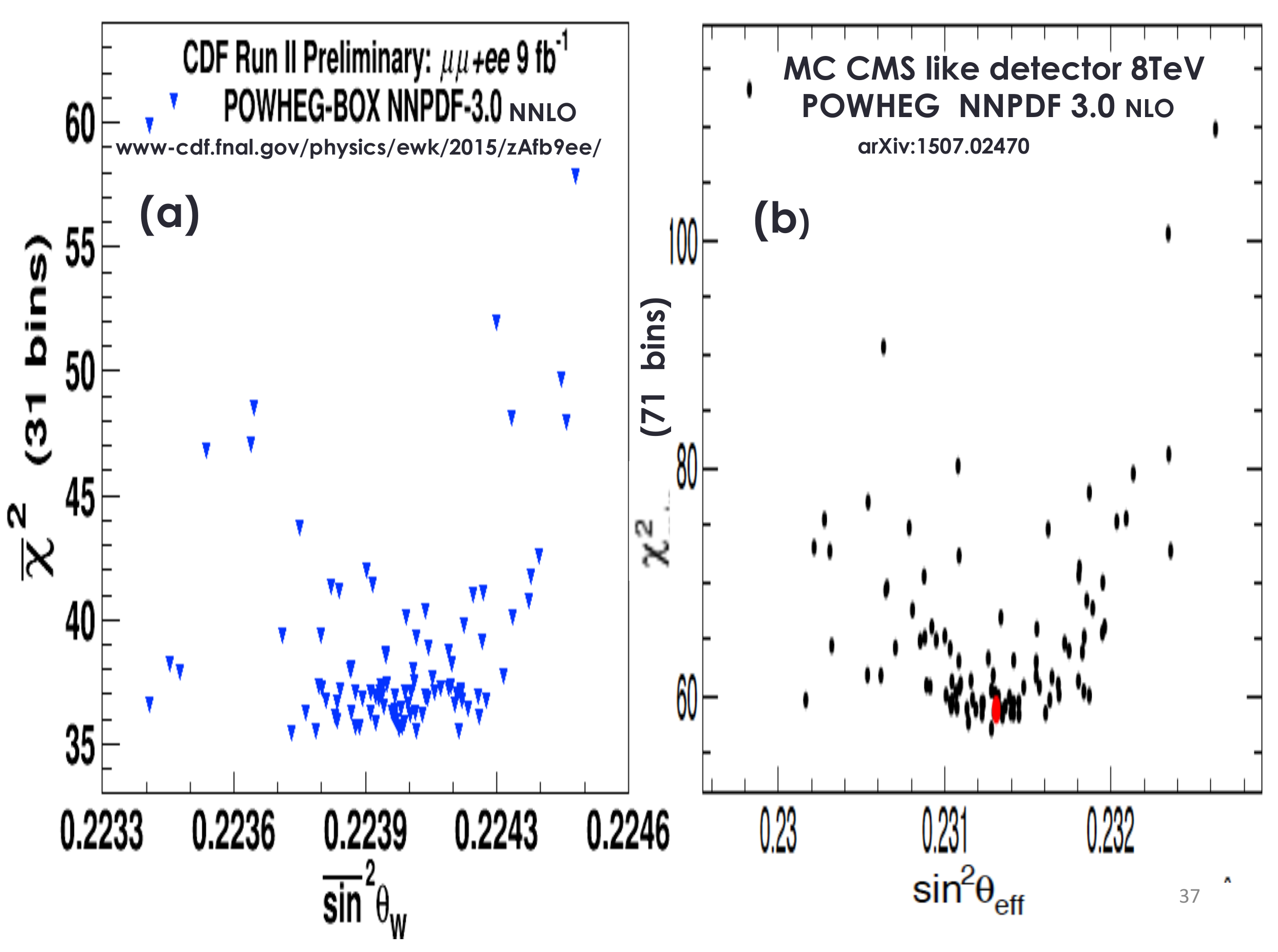}
\caption{ (a) CDF data: Best $\chi^2$ versus $\sin^2\theta_W$ (from ref. \cite{CDFweb}). (b) 
Best $\chi^2$ versus   $\sin^2\theta^{\rm lept}_{\rm eff}$(M$_Z$)  for MC simulation of a  CMS like detector with  15 fb$^{-1}$ at 8 TeV  (from ref.~\cite{pdf_error} (arXiv:1507.02470).}
\label{Fig4}
\end{figure}

 Fig.~\ref{Fig4}(a) shows the $\chi^2$ for the best fit value of $\sin^2\theta_{W}$  at CDF extracted using each of the 100 PDF replicas for the \textsc{nnpdf3.0} (\textsc{nnlo}) PDF set\cite{NNPDF}.   As shown in Fig.\ref{scale}(b) different values of $\sin^2\theta_{W}$ raise or lower $A_{\rm fb}$(M) for all values of  dilepton mass.  In contrast, as shown in Fig.\ref{scale}(a)  PDFs which raise the value of $A_{\rm fb}(M)$ for dilepton mass above the mass of the Z boson, reduce $A_{\rm fb}(M)$ below the mass of the Z bosons.  The sensitivity
 of $A_{\rm fb}(M)$ to $\sin^2\theta_{W}$ is very different from the sensitivity to PDFs. 
 Therefore, PDFs with a high 
 value of $\chi^2$ are less likely to be correct.  
 As shown in ref. \cite{pdf_error}, this information can be incorporated into the analysis by weighting the PDF replicas by $e^{-\chi^2/2}$. This reduces the weights of PDFs with large values of   $\chi^2$.  
 In addition to the measurements of $\sin^2\theta^{\rm lept}_{\rm eff}$(M$_Z$)
 and the on-shell $\sin^2\theta_{W}= 1- M_W^2/M_Z^2$,
  these  $A_{\rm fb}(M)$  constrained PDF weights can also  be used to reduce the PDF uncertainties in other Tevatron measurements such as the direct measurement of $M_W$.
\section{Results}
The Tevatron results with the full 9.4 fb$^{-1}$ sample are:
\begin{itemize}
\item D0:~~~ $\sin^2\theta^{\rm lept}_{\rm eff}$(M$_Z)$=0.23147$\pm$ 0.00043 (stat) $\pm$0.00008( syst)$\pm$0.00017 (\textsc{nnpdf2.3} \textsc{nlo} PDFs), \\$or  ~~~~~~~~~~~~~~~~~~~~$$\sin^2\theta^{\rm lept}_{\rm eff}$(M$_Z)^{D0}$~=~0.23147$\pm$ 0.00047
\item CDF: ~$\sin^2\theta^{\rm lept}_{\rm eff}$(M$_Z)$=0.23222$\pm$ 0.00042 (stat) $\pm$0.00008( syst)$\pm$0.00016 (\textsc{nnpdf3.0} \textsc{nnlo} PDFs), \\ $or~~~~~~~~~~~~~~~~~~~$$\sin^2\theta^{\rm lept}_{\rm eff}$(M$_Z)^{CDF}$ = 0.23222$\pm$ 0.00046
\item CDF:~~M$_W^{indirect}$ = 80.327$\pm$0.021(stat) $\pm$0.010(syst) GeV, 
 \\$or  ~~~~~~~~~~~~~~~~~~~~$M$_W^{indirect}$ = 80.327$\pm$0.023 GeV
\end{itemize}

\begin{figure}[h]
\centerline{\includegraphics[height=100mm,width=1.\linewidth]{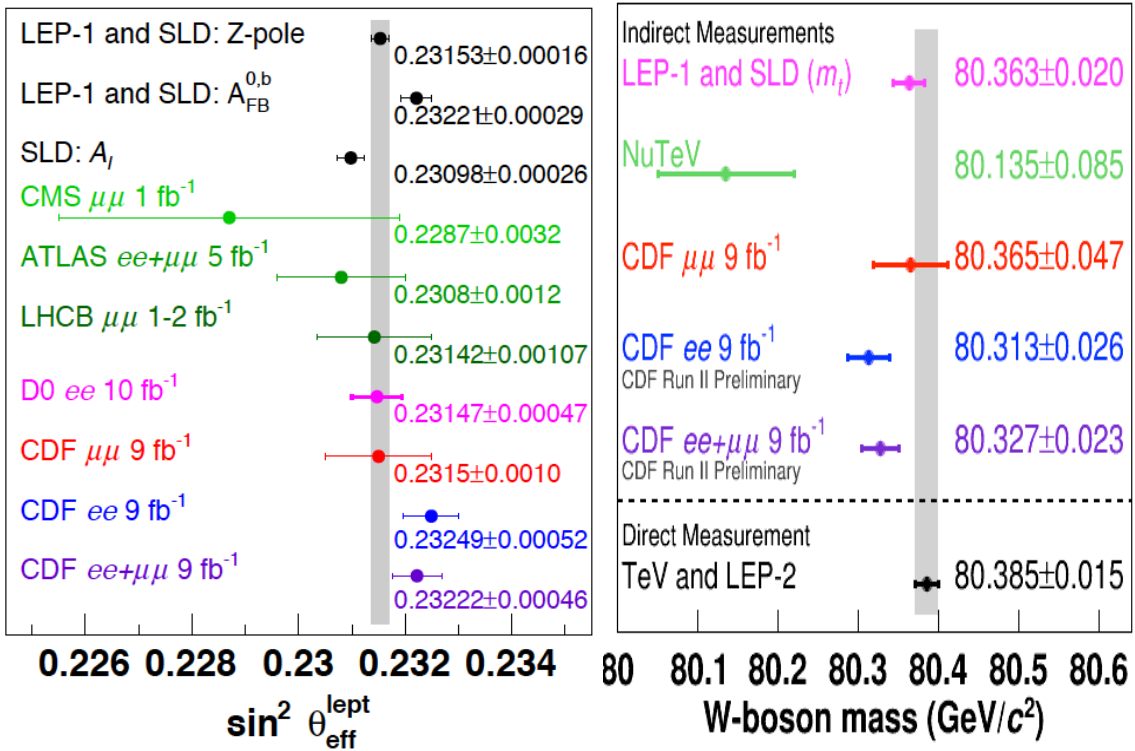}}
\caption{ Left panel: Comparison of $\sin^2\theta^{\rm lept}_{\rm eff}$(M)$_Z$ measurements.
 that includes the latest LHC results from CMS\cite{CMS_sw2}, ATLAS\cite{ATLAS_sw2} and   LHCb\cite{LHCb}.
The LEP-1+SLD Z-pole entry is the combination of their six Z-pole measurements. 
Right panel: $M_W$ measurements. All except for 'TeV and LEP-2' are indirect W-mass measurements that use the standard model (on-shell scheme). NuTeV is the Tevatron neutrino neutral current measurement\cite{nutev}. 
 }
\label{fig-sin2eff-mw}
\end{figure}

The left panel of Fig.\ref{fig-sin2eff-mw} shows a comparison of $\sin^2\theta^{\rm lept}_{\rm eff}$(M$_Z$)  measurement
from the Tevatron and other experiments, including the  latest LHC results from CMS\cite{CMS_sw2}, ATLAS\cite{ATLAS_sw2} and LHCb\cite{LHCb}.
The LEP-1+SLD Z-pole entry is the combination of their six Z-pole measurements. The right panel of Fig.\ref{fig-sin2eff-mw}
shows a comparison of CDF $M_W^{indirect}$ measurements to measurements by other experiments. 
 The  TeV and LEP-2 value is the world average of the direct measurements\cite{w2014} of $M_W$ ($M_W^{direct}=80.385\pm0.015$ GeV).  All the others  are indirect W-mass measurements that use the standard model (on-shell scheme).
  The indirect measurement labeled NuTeV\cite{nutev}  is the Tevatron neutrino neutral current measurement\cite{nutev}. The  indirect measurement
 labeled LEP1+SLD($M_t$)  is from standard model fits to all Z pole measurements\cite{LEPfinalZ}  in combination with the Tevatron
 top-quark mass measuremen\cite{top2014}.
%
%


\end{document}